# Hybrid approach for Image Encryption Using SCAN Patterns and Carrier Images

Panduranga H.T
Dept. of Electronics and Communication Engineering
Govt. Engineering College.
Hassan, Karnataka, India

Naveen Kumar S.K
Dept. of Electronics, PG Center
University of Mysore
Hassan, Karnataka, India

*Abstract*— We propose a hybrid technique for image encryption which employs the concept of carrier image and SCAN patterns generated by SCAN methodology. Although it involves existing method like SCAN methodology, the novelty of the work lies in hybridizing and carrier image creation for encryption. Here the carrier image is created with the help of alphanumeric keyword. Each alphanumeric key will be having a unique 8bit value generated by 4 out of 8-code. This newly generated carrier image is added with original image to obtain encrypted image. The scan methodology is applied to either original image or carrier image, after the addition of original image and carrier image to obtain highly distorted encrypted image. The resulting image is found to be more distorted in hybrid technique. By applying the reverse process we get the decrypted image.

I. INTRODUCTION

The development of internet provides a new way for spreading the digital information more conveniently. Information is an asset that has a value like any other asset. As an asset, information needs to be secured from the attacks. Because of the characteristic of digital images, some security problem comes out besides the extensive usage of these images. The importance of securing information/image has reached its highest levels in the recent years due to hacker attack and instruction of people's privacy.
Cryptography is popularly known as an art and science of secret writing. This enables us to store sensitive information or transmit it across insecure networks so that it cannot be read by anyone except the intended recipient. Image encryption has application in internet communication, video conferencing, telemedicine, distance education through video on demand, multimedia systems, military and satellite image processing.
S.R.M Prasanna and Y.V. Subba Rao have presented the method which employs magnitude and phase manipulation using carrier images [1]. The novelty of this paper explains the concept of carrier images for encryption purpose. Here, first apply the 2D-DFT to original image and randomly generated carrier images, then manipulation of phase data or magnitude data or both is done in the frequency demine, finally encrypted image is obtained by applying 2D-IDFT to manipulated images. Chao-Shen Chen and Rong-jain Chen proposed an image encryption and decryption algorithm based on SCAN methodology [2]. SCAN is a language-based two dimensional spatial-assessing methodology which can efficiently specify and generate a wide range of scanning paths. Here scanning path sequence fill the original image to generate the encrypted image. In [3] S.S Maniccam and N.G. Bourbakis have presented a method for image and video encryption, here first stage lossy video compression based on frame difference before the encryption. Here they say image encryption is performed by SCAN-based permutation of pixels and substitution rule which together form an iterated product cipher. Aloka Sinha and Kehar Singh proposed a new technique to encrypt an image for secure image transmission. Here the digital signature of the orthogonal image is added to the encoded version of the original image. The encoding of image is done using Bose-Chaudhuri Hocquenghm (BCH) code. The digital signature are created and verified by means of cryptography [4]. We implemented an encryption method which involves three steps; first step is constructing a size extended binary image using original image. In second step, applying the existing SCAN patterns to rearrange the pixels of extended binary image and finally in third step reconstruct the grayscale image to obtain encrypted version [5]. We also developed the concept of generating the carrier image with the help of unique code called as 4 out of 8-code, by adding the carrier image to original image we get the encrypted image [6].
The rest of the portion is as follows: section two gives idea behind the proposed work. Section three explains the brief overview of SCAN patterns. Section four explains the brief overview of carrier image creation using 4outof8 code. The proposed scheme for image encryption is described in Section five. Experimental observation and results are discussed in section six. Conclusion and future discussions are made in section seven.

II. IDEA BEHIND THE PROPOSED WORK

As we discussed in the introduction, there are several method for image encryption which deals in their own ideas. In few image encryption algorithms, encryption process depend only on the keywords, but in some other algorithms they use only carrier image for encryption. Anyhow, we have an idea to hybrid the existing algorithms to get a new path for encryption by taking the advantages of individual methods. Hence we come up with the concept of hybridizing SCAN pattern and Carrier image for Image encryption to get highly distorted Images.

## III. BRIEF OVERVIEW OF SCAN PATTERNS

A scanning of a two dimensional array is an order in which each element of the array is accessed exactly once. The SCAN is a formal language-based two dimensional spatial accessing methodology which can represent and generate a large number of wide varieties of scanning paths. SCAN language uses four basic scan patterns. They are continuous raster C, continuous diagonal D, continuous orthogonal O, and spiral S. Each basic pattern has eight transformations numbered from 0 to 7. For each basic scan pattern, the transformations 1, 3, 5, 7 are reverses of transformations 0, 2, 4, 6, respectively. The basic scan patterns and transformations are shown [2,3].

## IV. CARRIER IMAGE CREATION USING 4 OUT OF 8 CODE

Here we are defining a new code called 4 out of 8 code. This code is of 8 bit length with 4 number of one's and 4 number of zero's and we made one consideration that each nibble must have 2 number of ones and 2 number of zeros. We listed all 36 possible combinations of the 4 out of 8 code and each code is assigned to an alphanumeric character in table 1. Since 26 alphabets (capital letters or small letters) and 10 numerals forms to give 36 alphanumeric characters, this code is more suitable to assign a unique code to each alphanumeric character.

As we enter the different keywords, each keyword is taken and rearranged in a matrix form of size equal to the size of original image. If the length of the keyword is very small then the same keyword is repeated till the length is become equal to size of original image. By using luck up table of the alphanumeric character and 4 out of 8 code as shown in table1, a carrier image is created. Depending upon the keyword, carrier image is generated and used in the addition process to generate a encrypted image.

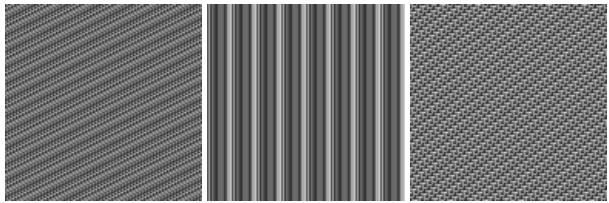

Figure1: Carrier image1, Carrier image2, Carrier image3.

Figure 1 shows the carrier images for the different keywords as fallows, key1='Iwant2EncryptThisImage', key2 = 'HybridApproch128z', key3='MyDateOfBirthIs21May1983'

| SL NO. | BIN | HEX | DEC | ALPHA NUMERIC |
|---|---|---|---|---|
| 1 | 00110011 | 33 | 51 | A,a |
| 2 | 00110101 | 35 | 53 | B,b |
| 3 | 00110110 | 36 | 54 | C,c |
| 4 | 00111001 | 39 | 57 | D,d |
| 5 | 00111010 | 3A | 58 | E,e |
| 6 | 00111100 | 3C | 60 | F,f |
| 7 | 01010011 | 53 | 83 | G,g |
| 8 | 01010101 | 55 | 85 | H,h |
| 9 | 01010110 | 56 | 86 | I,i |
| 10 | 01011001 | 59 | 89 | J,j |
| 11 | 01011010 | 5A | 90 | K,k |
| 12 | 01011100 | 5C | 92 | L,L |
| 13 | 01100011 | 63 | 99 | M,m |
| 14 | 01100101 | 65 | 101 | N,n |
| 15 | 01100110 | 66 | 102 | O,o |
| 16 | 01101001 | 69 | 105 | P,p |
| 17 | 01101010 | 6A | 106 | Q,q |
| 18 | 01101100 | 6C | 108 | R,r |
| 19 | 10010011 | 93 | 147 | SS |
| 20 | 10010101 | 95 | 149 | T,t |
| 21 | 10010110 | 96 | 150 | U,u |
| 22 | 10011001 | 99 | 153 | V,v |
| 23 | 10011010 | 9A | 154 | W,w |
| 24 | 10011100 | 9C | 156 | X,x |
| 25 | 10100011 | A3 | 163 | Y,y |
| 26 | 10100101 | A5 | 165 | Z,z |
| 27 | 10100110 | A6 | 166 | 0 |
| 28 | 10101001 | A9 | 169 | 1 |
| 29 | 10101010 | AA | 170 | 2 |
| 30 | 10101100 | AC | 172 | 3 |
| 31 | 11000011 | C3 | 195 | 4 |
| 32 | 11000101 | C5 | 197 | 5 |
| 33 | 11000110 | C6 | 198 | 6 |
| 34 | 11001001 | C9 | 201 | 7 |
| 35 | 11001010 | CA | 202 | 8 |
| 36 | 11001100 | CC | 204 | 9 |

Table 1: 36 possible combination of 4 out of 8 coed along with alphanumeric character.

## V. THE PROPOSED SCHEME FOR IMAGE ENCRYPTION

Encryption of an image can be done at different stage or in multiple stages and in multiple ways. If the encryption process is only in single stage then security is less as compare to multistage encryption. Figure 2 shows the block diagram of existing encryption approach using text as a key word and image as keyword. Encryption process may take different approaches, for example encryption may be using only SCAN method or only by Phase-Magnitude manipulation method or using only carrier image generated by 4 out of 8 code. Figure 3 shows the block diagram of proposed hybrid approach which contain multiple keywords, where key-1 is corresponding to create a carrier image and key-2, key-3, key-4 along with SCAN encryption process are optional.

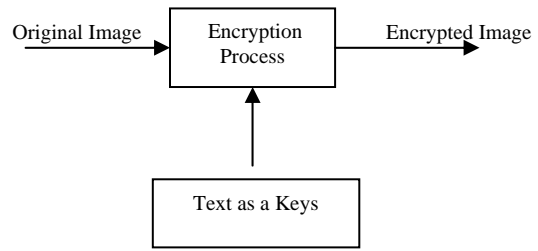

(a)

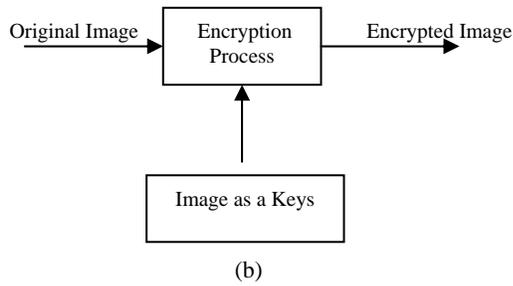

Figure 2. Block diagram of existing image encryption approach using (a) text as a key and (b) Image as a key

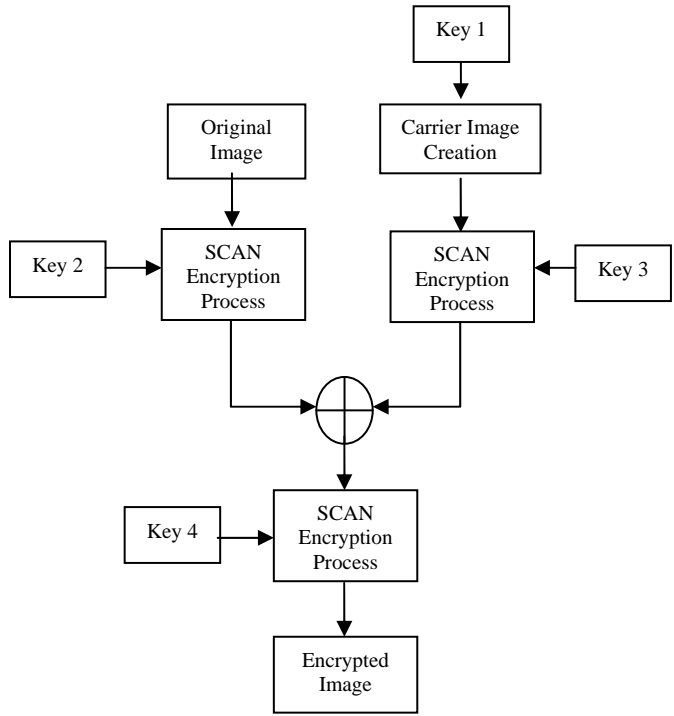

Figure 3. Block diagram of proposed image encryption approach using SCAN patterns and Carrier Image

## VI. RESULTS AND DISCUSSION

Here we have taken "lena.bmp" as a reference image to apply deferent encryption algorithm. Figure 4 shows the different encrypted image when only SCAN encryption process is done for different keys. Figure 5 shows the encrypted images if original image is added with the carrier images for different keywords i.e. key1 in the block diagram. Figure 6 shows encrypted images using SCAN encryption process at the input of original image and addition of carrier image. Figure 7 shows encrypted images using SCAN encryption process at different position along with the carrier image.

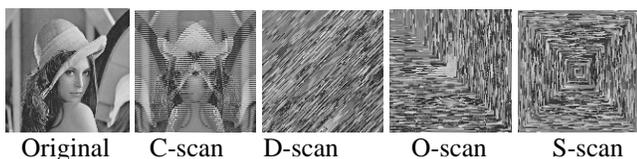

Original   C-scan   D-scan   O-scan   S-scan
Figure 4. Original image and encrypted image with different SCAN patterns

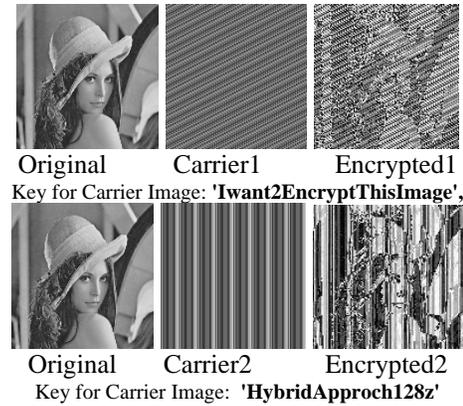

Original   Carrier1   Encrypted1
Key for Carrier Image: **'Iwant2EncryptThisImage'**,

Original   Carrier2   Encrypted2
Key for Carrier Image: **'HybridApproch128z'**
Figure 5. Original image, Carrier image and encrypted image for different keys at key1

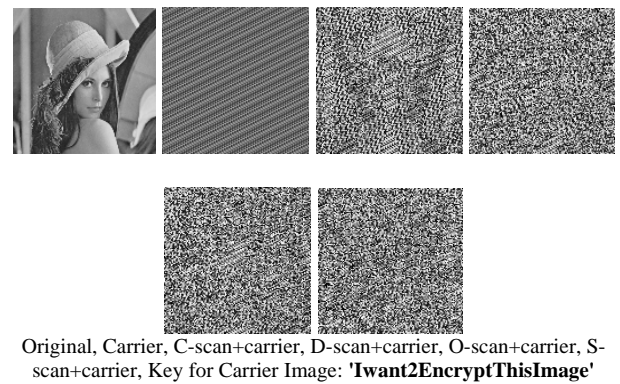

Original, Carrier, C-scan+carrier, D-scan+carrier, O-scan+carrier, S-scan+carrier, Key for Carrier Image: **'Iwant2EncryptThisImage'**

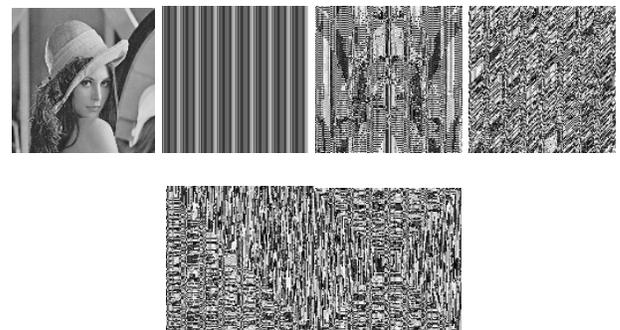

Original, Carrier, C-scan+carrier, D-scan+carrier, O-scan+carrier, S-scan+carrier, Key for Carrier Image: **'HybridApproch128z'**
Figure 6. Original image, Carrier image and encrypted image for different keys at key1 and SCAN Patterns

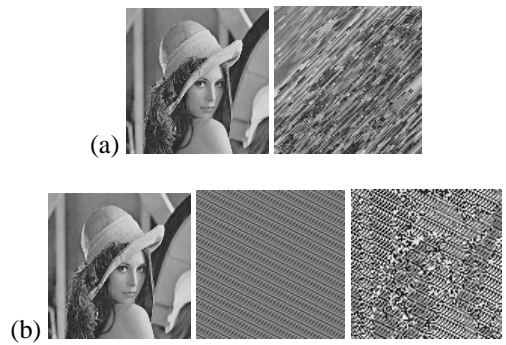

(a)

(b)

Panduranga H.T. et al. / (IJCSE) International Journal on Computer Science and Engineering
Vol. 02, No. 02, 2010, 297-300

## VII. CONCLUSION

From the experimental results we conclude that the proposed hybrid approach for image encryption gives very good result as compare to individual encryption process. For the sake of simplicity we used only few SCAN patterns and few carrier-keywords, but proposed algorithm also works for complex SCAN pattern and complex carrier-keywords. As the complexity increases the encrypted image is more distorted as compare to the result of single stage encryption. In future we want to develop an hybrid algorithm which uses more SCAN patterns, Carrier images along with some bit-manipulation techniques to provide more security to the images.

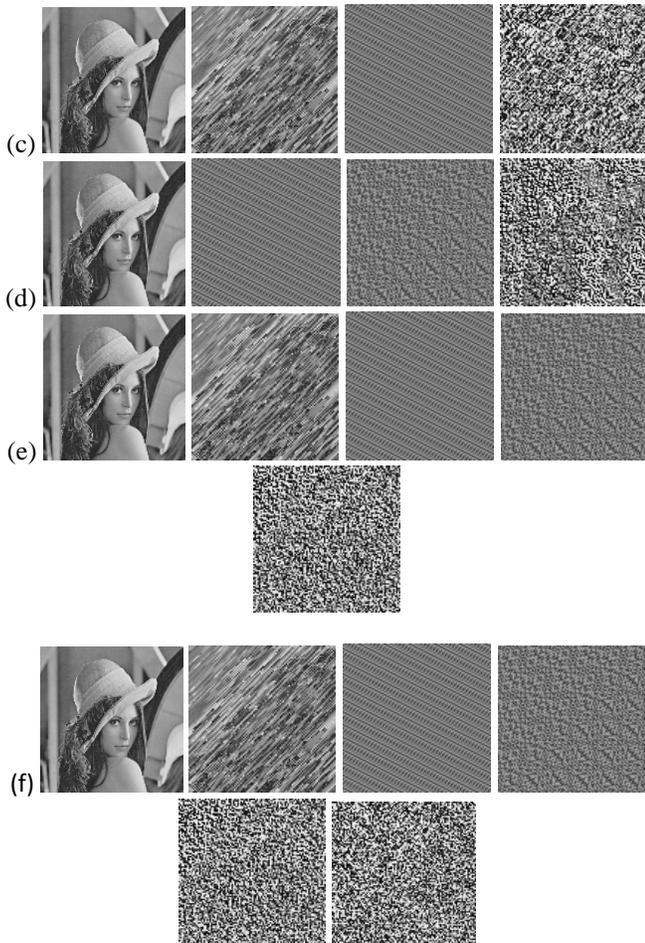

Key for Carrier Image: **'UniversityOfMysore'**

Figure 7. First image is the input image and last image is the Encrypted image in each row. (a)encrypted image = dscan (original_image),
(b) encrypted image = addition (original_image, carrier_image),
(c) encrypted image = addition (dscan (original_image), carrier_image),
(d) encrypted image = addition (original_image, dscan ( carrier_image))
(e) encrypted image =dscan ( addition (dscan(original_image), dscan ( carrier_image)))